# Factorial Unitary Representations of the Translational Group, Invariant Pure States and the Supersymmetric Graviton

J Moffat and C Wang

*Dept of Physics University of Aberdeen, UK*

**Abstract**

Unitary Representations corresponding to local shifts in reference frames are a key topic for progress in Quantum Gravity. The fibre bundle construct defined in our previous work in which quantum fields become liftings of; or sections through; a fibre bundle with base space curved space-time, continues to be the context for this paper. We investigate in more depth the subgroup *T* of the Poincare group consisting of translations of space-time as a gauge group of automorphisms. We define a representation of *T* as a group of automorphisms of the local fibre algebra **A(x)** which we assume to be isomorphic to a von Neumann algebra with trivial centre acting on a separable Hilbert space. Provided this group representation $\alpha$ is weakly measurable, then we have previously proved that it is also norm continuous, and is implemented by a norm, hence weakly and strongly, continuous unitary representation, by unitaries in the algebra. We discuss the dependence of this key result on the strange properties of Stonean spaces and extend our exploitation of Mackey Theory to give a more straightforward group theoretic proof. From such unitary representations a minimal form of Supersymmetry naturally emerges, which predicts the existence of both the mass zero graviton, and its gravitino partner.

**Introduction**

Our focus is on the local fibre algebra **A(x)** which we assume to be isomorphic to a von Neumann algebra with trivial centre acting on a separable Hilbert space. We also consider the subgroup *T* of the Poincare group consisting of translations of space-time as a gauge group of automorphisms of **A(x)**. These generate the local diffeomorphisms of General Relativity.

Consider then the automorphic representation $g \to \alpha_g$ of the translation subgroup $T$. If we make the minimal assumption that this representation is weakly measurable; ie the mapping $g \to \langle \alpha_g(A)x, y \rangle$: is Haar-measurable for all relevant values of *A, x* and *y*; then the argument of [1] shows that the mapping $g \to \alpha_g$ is norm continuous. Since the translational group is both abelian and connected, it follows that each $\alpha_g$ is an inner automorphism of **A** and the corresponding unitary $W_g$ has a spectrum contained in the positive half plane. In fact we have;

$$\sigma(W_g) \subset \{z; \operatorname{Re} z \geq \beta_g\} \text{ where } \beta_g = \frac{1}{2}\left(4 - \|\alpha_g - i\|^2\right)^{\frac{1}{2}}$$

Moreover if **S** denotes the von Neumann subalgebra generated by $\{W_g; g \in T\}$, the set of unitaries $\{W_g; g \in T\}$ is actually a commuting set within **A(x).** Thus **S** is a commutative subalgebra and contains the identity I of **A(x)**, since if *id* is the group identity then $I = W_{id}$. **A(x)** is a factor, **S** thus contains the centre Z of **A(x)**. Such a commutative quantum operator algebra is equivalent to the set of continuous functions on a compact space and this equivalence arises through the Gelfand transform;

$$A \to \hat{A} \text{ with } \hat{A}(\rho) = \rho(A) \text{ and } \rho \text{ a continuous complex valued homomorphism.}$$

The carrier space $\Phi_S$ of **S** is the set of all continuous complex valued homomorphisms on **S**. and is thus topologically a Stonean space, as is $\Phi_Z$ and the restriction map $\pi: \Phi_S \to \Phi_Z$ is a continuous surjection**.** We then have, as we will prove, a lifting

$$f: \Phi_Z \to \Phi_S \text{ with } f(\rho)|_Z = \pi \circ f(\rho) = \rho \text{ for } \rho \in \Phi_Z.$$

For each $g \in T$, we can now define the Gelfand transform of a unitary $U_g$ acting on the carrier space $\Phi_S$ of **S** as follows;

$$\hat{U}_g(\rho) = \overline{\hat{W}_g}(f(\rho|_Z))\hat{W}_g(\rho) = \overline{\hat{W}_g}(f \circ \pi(\rho))\hat{W}_g(\rho), \quad \rho \in \Phi_S$$

If we define the equivalence $\rho \approx \rho' \Leftrightarrow \rho|_Z = \rho'|_Z$ then by the extended form of the Stone-Weierstrass theorem [2] the centre Z corresponds to those elements of $\Phi_S$ constant on each

equivalence class, Applying this to the Gelfand transform $\hat{V}_g(\rho) = \overline{\hat{W}_g}(f(\rho|_Z))$ it follows [3]
that $\hat{V}_g$ corresponds to an element $V_g$ of the centre. Since **A(x)** is a factor, this means that
$V_g = \nu(g)I$ with $\nu(g)$ a complex number defining a coboundary, and that for each
$g \in T$; $U_g$ implements $\alpha_g$. In addition, the mapping $g \to U_g$ is a group homomorphism for if
we set;
$R_{g,h} = U_g U_h U_{gh}^*$ then for an operator A, $R_{g,h} A R_{g,h}^* = U_g U_h U_{gh}^* A U_{gh} U_h^* U_g^* = \alpha_g \alpha_h \alpha_{gh}^{-1}(A) = A$
Hence $R_{g,h}$ is a unitary in the centre. In fact $R_{g,h} = \lambda(g,h)I$ where $\lambda(g,h)$ is a 2-cocycle.
The fact that the lifting $f : \Phi_Z \to \Phi_S$ has the property $\pi \circ f(\rho) = f(\rho)|_Z = \rho$ for $\rho \in \Phi_Z$ means
that $\hat{R}_{g,h}(\rho) = \hat{R}_{g,h}(f(\rho))$ since the domain of $\hat{R}_{g,h}$ is $\Phi_Z$. But then we have that;
$\hat{U}_g(f(\rho)) = 1 \quad \forall g \in G$, thus $\hat{R}_{g,h}(\rho) = 1 \quad \forall \rho \in \Phi_Z$; the 2-cocycle is trivial.
Therefore $R_{g,h} = I \quad \forall g, h \in G$ and $g \to U_g$ is a group representation by unitaries
in the fibre algebra, which turns out to be norm continous, as previously shown [1, 7].
In summary, the lifting follows from the Axiom of Choice applied to the strange topological
properties of Stonean spaces. We can set it in the context of lifting from a totally disconnected,
compact Hausdorff base space S into a containing fibre bundle K having a Stonean topology.

*Properties of the Projection onto the Base Space B of a Stonean Fibre Bundle K.*

We prove first the subsidiary fact that the projection $\pi$ is an open mapping if and only if for
each non-trivial open subset *E* of *K*, *π(E)* is not a nowhere dense subset of the base space *B*.

One way is trivial for if $\pi$ is an open mapping then *π(E)* is a non-trivial open set so cannot be
nowhere dense.
Conversely, the topology of the fibre bundle K is compact and totally disconnected, with a basis
of 'clopen' sets (i.e. sets which are both closed and open). Thus every open set is a union of
such clopen sets, and it suffices to show that if V is clopen in K, then *π(V)* is open in B. Since V
is clopen, it is a closed and thus compact subset of K, and $\pi$ is continuous, thus *π (V)* is
compact. If we define $Y = \pi(V) \setminus \text{int } \pi(V)$; this a closed set with empty interior thus Y is a nowhere dense set, and is the image of an open set; $Y = \pi\left(V \setminus \pi^{-1}(\text{int } \pi(V))\right)$, using the fact that the projection mapping π is continuous and surjective. It follows that $Y$ is void and $\pi(V) = \text{int } \pi(V)$. Thus $\pi(V)$ is open and $\pi$ is an open mapping.

*Lifting from the Base Space B*

Consider now the set $\Phi$ of all compact subsets S of K such that $\pi(S) = B$. Then $\Phi$ is a non-void, partially ordered by set inclusion, and every decreasing chain has a lower bound. It follows from Zorn's Lemma that $\Phi$ has a smallest element K(0). We show that;

- $\pi|_{K(0)}$ is an open mapping ;
- $\pi|_{K(0)}$ is injective

This will prove the result for the lifting we seek is then $f = \left(\pi|_{K(0)}\right)^{-1}$, the Axiom of Choice selecting out K(0) as a unique minimal section through the fibre bundle K.

$\underline{\pi|_{K(0)} \text{ is an open mapping} ;}$

Consider then V to be a non-trivial open set in K(0). By definition, $\pi|_{K(0)}$ is surjective, thus; $B \setminus \pi(V) \subset \pi(K(0) \setminus V)$. Now $K(0) \setminus V$ is a closed thus compact subset of K, and $\pi|_{K(0)}$ is continuous, thus $\pi(K(0) \setminus V)$ is compact. It follows that $\overline{(B \setminus \pi(V))} \subset \pi(K(0) \setminus V)$. If $\pi(V)$ is nowhere dense, then;

$$B = B \setminus \text{int}\left(\overline{\pi(V)}\right) \subset B \setminus \overline{\pi(V)} \subset \pi(K(0) \setminus V), \text{ a closed, compact set.}$$

This is a contradiction due to minimality of the set K(0). Thus $\pi(V)$ cannot be a nowhere dense set. From our earlier discussion, this is enough to show that $\pi|_{K(0)}$ is an open mapping.

$\underline{\pi|_{K(0)} \text{ is injective} ;}$

Assume that $\pi|_{K(0)}$ is not injective, then $\exists x_1, x_2; x_1 \neq x_2 \in K(0)$ with $\pi(x_1) = \pi(x_2)$.

Since K(0) is a Hausdorff topological space with a basis of clopen sets, there is a clopen subset V of K(0) containing $x_1$ but not $x_2$. Then;

$\pi(V)$ is also clopen and $\pi^{-1}(\pi(V)) \setminus V \subset K(0)$, and $\pi(x_2) = \pi(x_1) \in \pi(V)$, thus $x_2 \in \pi^{-1}(\pi(V)) \setminus V$

. It follows that $W = \pi^{-1}(\pi(V)) \setminus V$ is a non-trivial open set, and W and V are disjoint, with

$B \setminus \pi(V) \subset \pi(K(0) \setminus W)$ and $\pi(V) \subset \pi(K(0) \setminus W) \Rightarrow \pi(K(0) \setminus W) = B$

This again contradicts the minimality of K(0). Thus $\pi|_{K(0)}$ is a continuous bijection, and the required lifting is given by $f = (\pi|_{K(0)})^{-1}$.

**Mackey Theory and a Group Theoretic Proof**

For a direct group theoretic proof, avoiding the intricacies of Stonean topology, we have devised the following argument which builds on the discussion of Mackey Theory applied to group extensions and short exact sequences in our previous work [1, 4]. To prepare the ground, we need the following seemingly little known result.

Define a character acting on a locally compact abelian group G to be a continuous group homomorphism from G to the unit circle. If H is a closed subgroup of G and $\rho$ is a character of H, then $\rho$ extends to a character on G.

Since this useful result plays an important part in our argument, we provide a simple proof as follows, based on Mackey's general approach.

Let $\hat{G}$ denote the group of characters on G under pointwise multiplication. Pontryagin [5] shows that we can impose a locally compact topology on $\hat{G}$ relative to which it is a locally compact abelian group. Define;

$$\omega(g)(\rho) = \rho(g) \quad \forall g \in G, \rho \in \hat{G}$$

Then the mapping ω is a continuous isomorphism between G and $\hat{\hat{G}}$.

Let $L = \{\rho \in \hat{G}; \rho(H) = 1\}$ and $S =$ the character group of the quotient group $\frac{G}{H}$
then $L = S$ for if $\beta \in S \Leftrightarrow \alpha: g \to \beta(Hg) \Leftrightarrow \alpha(h) = 1 \quad \forall h \in H \Leftrightarrow \alpha \in L$

If we move this up a level and now define $S$ as the character group of the quotient group $\frac{\hat{G}}{L}$ then $S$ is the subset of $G = \hat{\hat{G}}$ which is constant on $L$ but up to isomorphism, $H = \hat{\hat{H}}$ is constant on $L$ thus in this case $S=H$.

Finally we can conclude that, up to isomorphism, $\hat{H} = \hat{S} = \frac{\hat{G}}{L}$ since S is the character group of the quotient group $\frac{\hat{G}}{L}$.

Let $\varphi$ denote the isomorphism: $\hat{H} \to \frac{\hat{G}}{L}$. If ψ denotes the restriction mapping

$\rho \to \rho|_H : \hat{G} \to \hat{H}$, then the kernel of ψ is L. Thus $\psi = \varphi^{-1}$. If Ω is the quotient mapping:

$\hat{G} \to \frac{\hat{G}}{L}$ then we have;

For $\xi \in \hat{H}$ the mapping $\Omega^{-1} \circ \varphi : \hat{H} \to \frac{\hat{G}}{L} \to \hat{G}$ thus $\rho = \Omega^{-1} \circ \varphi(\xi) \in \hat{G}$

and then $\rho|_{\hat{H}} \equiv \psi(\rho) \equiv \psi(\rho L) \equiv \psi \circ \Omega(\rho) = \xi$ since $\varphi^{-1} = \psi$

and $\xi \to \rho$ is the require extension to the character group G of the character group of the subgroup H.

By exploiting Mackey Theory directly, we now aim to prove the following key result by an alternative method to that used in our previous paper [1] in order to help stimulate a broader interest in such group theoretic approaches, which extend the work of Wigner.

*Define $\alpha : g \to \alpha_g$ to be a representation of the translational subgroup T of the Poincare group as a gauge group of automorphisms of the fibre algebra A(x). Provided this group representation $\alpha$ is weakly measurable, then it turns out that it is also norm, weakly and strongly continuous, and is implemented by a norm, weakly and strongly continuous unitary representation by unitaries in the fibre algebra.*

We showed in our previous paper [1] that, with these assumptions, the mapping $\alpha : g \to \alpha_g$ is norm continuous and is thus implemented by a set $g \to W_g$ of commuting unitaries in the fibre algebra. Since $\|\alpha_g - i\| \geq \{\|\alpha_g(T)x - Tx\|; \ \|T\| \leq 1 \|x\| \leq 1\}$ it follows that the norm continuous mapping $\alpha : g \to \alpha_g$ is strongly continuous. Our foundational work at [6] then shows that we may choose the unitaries in a way that the mapping $g \to W_g$ is a borel mapping from T to the unitary group of **A(x)** with the weak operator topology.

Let $N = Ker(\alpha) = \{g \in T; \alpha_g = \text{the identity}\}$. Replacing $T$ by $T/N$ we may assume that;

$$\alpha_g = \alpha_h \text{ implies } g = h. \ldots\ldots\ldots(1).$$

We note that the unitary group of the centre of **A(x)** is isomorphic to the unit circle **O** in the complex plane. It is also easy to see that $\Gamma = \{\lambda W_g; \lambda \in \mathbf{O}, \text{ and } g \in T\}$ is an abelian subgroup of the unitary group of **A(x)**.

Define the mapping $\gamma(\lambda) = \lambda I$ which maps the unit circle **O** into the subgroup $\Gamma$, and the mapping $\eta(\lambda W_g) = g$ which maps the group $\Gamma$ to the translation group *T*. Note that the mapping $\eta$ is well defined here due to (1) above. Then the sequence;

$$0 \to \mathbf{O} \xrightarrow{\gamma} \Gamma \xrightarrow{\eta} T \to 0$$

is short exact. Thus $\Gamma$ is an extension of **O** by *T*. We can identify the group $\Gamma$ with the extension $\mathbf{T}\eta'T$ where; $\eta'(\lambda, g) = \eta(\lambda W_g) = g$ via the mapping; $(\lambda, g) \to \lambda W_g$.

Let the mapping *J* denote the identification $\mathbf{O} \times T \leftrightarrow \Gamma \leftrightarrow \mathbf{O}\eta'T$ then *J* is a borel mapping. Thus we see again that the mapping $W: g \to W_g$ is a borel measurable mapping from the translation subgroup *T* into the group $\Gamma$.

The group *T* is a separable locally compact abelian group and is a standard borel space. The unitary group of **A(x)** is also a standard borel space, thus $W^{-1}: W_g \to g$ is a borel mapping and $\eta'$ is a borel system of factors for $\mathbf{T}\eta'T$. There is thus, as discussed in [1, 6], a locally compact topology on the group $\Gamma$ relative to which it is a separable, locally compact abelian group. $\Gamma$ also contains the unit circle **O** as a closed subgroup.

With this Mackey topology on $\Gamma$, we can now deduce that the identity map $e^{i\theta} \to e^{i\theta}: \mathbf{O} \to \mathbf{O}$ is a character on **O** and thus lifts to a character $\sigma: \Gamma \to \mathbf{O}$.

For each unitary $W_g; g \in T$ define $U_g = \sigma(W_g)^* W_g$. Then;

$U_g$ implements the automorphism $\tau_g$ and, since $\sigma|_\mathbf{O}$ is the identity;

$$\sigma(U_g) = \sigma(W_g)^* \sigma(W_g) = 1 \ \forall g \in T.$$

Now if :

$g, h \in T$ then $U_g U_h = \lambda(g,h) U_{gh}$ for a 2-cocycle $\lambda(g,h) \in \mathbf{O}$

Since $U_g \in \Gamma \ \forall g$, by definition of $\Gamma$, we can apply the group homomorphism $\sigma$ to both sides of this relationship. Noting again that $\sigma|_O$ is the identity, this implies that all such 2-cocycles are trivial, and the mapping $g \to U_g$ is a unitary group representation.

From the proof above we can identify the group $\Gamma$ with the borel system of factors $\mathbf{O}\eta'T$

The mapping $W: g \to W_g$ is a borel measurable mapping from the translation subgroup $T$ into the group $\Gamma$. Hence the mapping $g \to \langle x, W_g x \rangle$ is a measurable mapping for any $x$ in the Hilbert space $H$. In addition, the character $\sigma: \Gamma \to \mathbf{O}$ is a continuous mapping, thus $\sigma \circ W: g \to \sigma(W_g)$ is borel. Combining these together it follows that the mapping $g \to \langle x, U_g x \rangle = \sigma(W_g)^* \langle x, W_g x \rangle$ is borel measurable and is in fact norm continuous due as before to the spectral properties of the $\{W_g; g \in G\}$ which imply that $\|U_g - I\| \leq 2\sqrt{2 - 2\beta_g}$ and $\beta_g \to 1$ as $g \to e$.

**Pure *T*-invariant Quantum States**

We continue to assume as context that space-time is non-commutative at some energy level such as the Planck regime, with algebraic structure at each event point **x** of space-time, forming the fibre algebra **A(x)**. This structure then corresponds to the single fibre of a principal fibre bundle. A gauge group of automorphisms corresponding to the translation subgroup T of the Poincare group acts on each fibre algebra locally, while a section through this bundle is then a quantum field of the form $\{A(x); x \in M\}$ with *M* the underlying space-time manifold. In addition, we assume a local algebra *O(D)* corresponding to the algebra of sections of such a principal fibre bundle with base space a finite and bounded subset of space-time, *D*. The algebraic operations of addition and multiplication are assumed defined fibrewise for this algebra of sections.

We define a *separating family* of *T*-invariant quantum states to be a finite or countable subset *S* of the state space such that the positive kernel of *S* is zero; i.e. given an observable *A* in the positive subset of the fibre algebra **A(x)**,

$$f_j(A) = 0 \; \forall f_j \in S \Rightarrow A = 0$$

For a general observable *A* we assume that $f(A^*A) = 0 \Rightarrow A = 0$. This reflects the classical case; if a tensor or the difference between two tensors of the same covariant and contravariant class is equal to zero for all local inertial reference frames, then it is zero for all curvilinear reference frames, by the covariance assumptions of General Relativity.

Since the Hilbert space F(x) on which the fibre algebra **A(x)** acts is separable, by definition, there is a countable dense subset $x(n)$. Defining $f = \sum_{1}^{\infty} \alpha_{x(n)} \omega_{x(n)}; \lim_{n \to \infty} \sum_{1}^{n} \alpha_{x(j)} = 1$; clearly $f$ is a separating state, and then each element of the weak*-closed convex hull $\Delta = \overline{co}\{f \bullet \alpha_g; g \in T\}$ is also separating, for if;

$$\sum_j \lambda(j) f \circ \alpha(g_j)(A^*A) = 0 \Rightarrow \exists g_j \in T; f \circ \alpha(g_j)(A^*A) = 0 \Rightarrow \alpha(g_j)(A^*A) = 0,$$

since $f$ is separating.

Applying the automorphism $\alpha(g_j^{-1}) \Rightarrow A^*A = 0$. Additionally, if $f_n(A^*A) \to g(A^*A)$ and $f_n(A^*A) = 0$ then $g(A^*A) = 0$. Thus every element of $\Delta$ is separating.

From our previous work [1,4], $\Delta$ contains an invariant state. Thus there is a separating invariant quantum state. This has potential structural implications for the Wigner-Dirac representation theory of the underlying algebra A(x). For now we note that applying the Kovacs-Szucs Theorem [7]; if $f$ is a separating normal state, and if K(A) is the closed convex hull of the orbit of an observable A in the ultraweak operator topology; $K(A) = \overline{co}\{\alpha_g(A); g \in T\}$, then K(A) contains exactly one invariant element, which we denote as Π(A). The mapping Π: A ⟶ Π(A)) is a projection of norm 1, constant on each orbit $\{\alpha_g(A); g \in G\}$.

If $f$ is a separating T-invariant quantum state, let $\pi$ be the GNS representation then clearly $\pi$ is an algebraic isomorphism since the kernel of $f$ is {0}. Let $\{U_g; g \in G\}$ be the Segal unitaries associated with $f$; then the mapping $g \to U_g$ is a unitary representation implementing $\alpha : g \to \alpha_g$. If the mapping $\alpha : g \to \alpha_g$ is weakly measurable in the GNS representation then by our previous results [1] it is norm continuous and we can further assume [1, 3, 6] that there is a unitary mapping $g \to U_g$ implementing $\alpha : g \to \alpha_g$ which is norm continuous with $U_g \in \pi(\mathbf{A(x)}) \, \forall g \in T$.

In previous work [4] we characterised these invariant states in terms of their ergodic action focusing on density matrix states and their von Neumann entropy. From inspection, we can generalise many of these previous results to any (not necessarily normal) quantum state: see final section of the paper. We have, from that analysis; that the automorphic representation $\alpha : g \to \alpha_g$ of $T$ acts ergodically if and only if $\pi(A(x)) \cap \{U_g ; g \in T\}'$ is trivial, containing only the projections 0 and I and thus consisting of the set of complex multiples of $I$.

We have, from equation (3) of the final section;

$$J\left\{\pi(A(x)) \cap \{U_g ; g \in T\}'\right\} J = \pi(A(x))' \cap \{U_g ; g \in T\}' \ .$$

Since $J^2 = 1$, it follows that the representation $\alpha : g \to \alpha_g$ of $T$ acts ergodically if and only if $\pi(A(x))' \cap \{U_g ; g \in T\}'$ is also trivial. But for this case we have also shown that $U_g \in \pi(A(x)) \ \forall g \in T$ and thus if E is a projection in $\pi(A(x))'$ then clearly

$$E \in \pi(A(x))' \cap \{U_g ; g \in T\}' = \{\lambda I\} \ .$$

The GNS representation is thus irreducible in the sense of Murray-von Neumann, corresponding to *f* being a pure quantum state.

## An Example Application; The Supersymmetric Graviton

The Poincare group is a locally compact Lie group with 10 generators, and the translational group is an abelian subgroup generated by the energy-momentum 4-vector $P_\mu$. This has the property that its square $P^2 = P_\mu P^\mu = E^2$ lies in the centre of the Lie algebra. If we consider the energy-momentum vector in normalized units *(c=1)* then $P^2$ thus has the form $P^2 = m^2 I$, where *m* is the mass-energy of the corresponding particle. In other words, a factorial representation of the Translational group corresponds to a particle with fixed mass *m* and undetermined spin. We can consider two cases;

(a). $P^2 = m^2 . I ; m^2 \neq 0$ corresponding to a multiplet of particles each of the same positive mass but with different spin values;

(b). $P^2 = 0.I$. This factorial representation corresponds to a massless particle such as a photon or a graviton, with a Supersymmetric massless fermion partner.

For either case we need to add an additional element, normally denoted $Q_\alpha$, to the Lie algebra, to represent the spread of spin values. In any representation, these are all linear operators, including the identity operator *I*, and thus form an algebra of such operators. Such a factorial representation corresponding to a set of particles, must contain equal numbers of bosons and fermions [8]. With certain assumptions, such a representation where the centre of the algebra is non-trivial can be decomposed into a direct integral of factorial representations, as discussed in [6].

We can develop the local, linear representation of these operators, which is a faithful representation of the Superspace Lie Algebra [8] by adding a pair of Grassmann variables to the algebraic formulation. We can then generate a standard Lie algebra while mixing commutators anticommutators. For example, if $\xi$ and $\bar{\xi}$ have the Grassmann property so that $\xi\bar{\xi} = -\bar{\xi}\xi$, then; $\left[\xi Q_A, \bar{\xi}\bar{Q}_{\dot{B}}\right] = \left(\xi Q_A \bar{\xi}\bar{Q}_{\dot{B}} - \bar{\xi}\bar{Q}_{\dot{B}}\xi Q_A\right) = \xi\bar{\xi}\left(Q_A\bar{Q}_{\dot{B}} + \bar{Q}_{\dot{B}}Q_A\right) = \xi\bar{\xi}\{Q_A, \bar{Q}_{\dot{B}}\}$.

Note that we also have to assume that these Grassmann variables commute with the operators *Q*. A typical element of the corresponding Lie group *G* is then of the form;

$$G(x^\mu, \xi, \bar{\xi}) = \exp i(\xi Q + \bar{\xi}\bar{Q} - x^\mu P_\mu)$$

Here, $x^\mu$ is a point in flat space-time, thus we can think of the Grassmann variables as a vector at the point $x^\mu$. With this structure, the Superspace is a vector bundle and the locally flat group multiplication structure is of the following form;

$G(x^\mu, \theta, \bar{\theta})G(a^\nu, \xi, \bar{\xi}) = G(x^\mu + a^\nu - i\xi\sigma^\mu\bar{\theta} + i\theta\sigma^\mu\bar{\xi}, \theta + \xi, \bar{\theta} + \bar{\xi})$. This follows from the Grassmannian properties.

From our previous paper [1] we noted that the assumption of local flatness of the space-time manifold is equivalent to the generators $P_\mu$ ($\mu = 0, 1, 2, 3$) commuting and thus corresponds to a factorial representation of the translational group by commuting unitaries. It follows that this factorial representation is norm continuous and is implemented by a norm continuous unitary representation. We can interpret this group product as shifting the locus of the Grassmann vector in space-time from $x^\mu$ to $x^\mu + a^\mu - i\xi\sigma^\mu\bar{\theta} + i\theta\sigma^\mu\bar{\xi}$ together with additive change to the Grassmann vector at this point. If this shift is infinitesimal, then, as in normal Lie group theory, we can consider the tangent plane around the group element $G(a^\mu, \xi, \bar{\xi})$, giving the following local representation of the Lie group generators on the tangent plane to the Riemannian space-time manifold at the event point $(a^\nu)$:

$$P_\mu(a^\nu) = i\frac{\partial}{\partial x^\mu}\Big|_{(a^\nu)} = i\partial_\mu\Big|_{(a^\nu)}$$

$$iQ_A(a^\nu) = \frac{\partial}{\partial \theta_A}\Big|_{(a^\nu)} - i\sigma^\mu\bar{\theta}\frac{\partial}{\partial x^\mu}\Big|_{(a^\nu)}$$

$$i\bar{Q}_{\dot{A}}(a^\nu) = -\frac{\partial}{\partial\bar{\theta}_{\dot{A}}}\Big|_{(a^\nu)} + i\theta\sigma^\mu\frac{\partial}{\partial x^\mu}\Big|_{(a^\nu)}$$

In this form the generators all satisfy the algebraic relationships of the Supersymmetric extension of the local translation Lie algebra. Thus this representation is locally an algebraic isomorphism onto the curved Riemannian manifold $\mathcal{M}$ and the piecewise local representations;

$$P_\mu(a^\nu) = i\frac{\partial}{\partial x^\mu}\Big|_{(a^\nu)} = i\partial_\mu\Big|_{(a^\nu)}$$

are the generators of the local relativistic diffeomorphisms around the event points of $\mathcal{M}$. The manifold is assumed smoothly differentiable; we focus on the flat tangent plane and assume Dirac relativistic spinor theory applies. We therefore assume that this operator representation acts on a 4 dimensional Hilbert space $H$ of (spinor) wave functions; and we denote by $\psi$ an element of the Hilbert space.

Within these assumptions we have the Dirac matrix $\gamma^5 = i\gamma^0\gamma^1\gamma^2\gamma^3$, and we can express this state $\psi$ in the form $\psi = \frac{1}{2}(I_{4x4} - \gamma^5)\psi + \frac{1}{2}(I_{4x4} + \gamma^5)\psi$. We define these two terms as the left landed and right handed spinor components of the state vector, thus we define;

$$\psi_L = \frac{1}{2}(I_{4x4} - \gamma^5)\psi \text{ and } \psi_R = \frac{1}{2}(I_{4x4} + \gamma^5)\psi$$

So that $\psi = \psi_L + \psi_R$.

If we define the Dirac adjoint function $\bar{\psi} = \psi^+\gamma^0$ where $\psi^+ = (\psi^*)^T$ then taking complex conjugates of both sides of the massless Dirac equation followed by transposition yields the identity:

$$(\psi^*)^T(-i(\gamma^{\mu*})^T(\partial_\mu + ieA_\mu) = 0$$

Exploiting the fact that $(\gamma^0)^2 = I_{4x4}$ leads to the equation:

$$-i\gamma^{\mu T}(\partial_\mu + ieA_\mu)\bar{\psi}^T = 0$$

The matrices $(-\gamma^{\mu T})$ also satisfy the Clifford algebra relations and there is a 4x4 non-singular matrix $C$ such that $C^{-1}\gamma^\mu C = -\gamma^{\mu T}$. We can thus define the charge conjugate spinor

$$\psi^c = C\bar{\psi}^T.$$

In the Weyl representation we take;

$$C = i\gamma^0\gamma^2 = i\begin{pmatrix} -\sigma^2 & 0 \\ 0 & \sigma^2 \end{pmatrix} \text{ where } \sigma^2 \text{ is the second Pauli matrix } \begin{pmatrix} 0 & -i \\ i & 0 \end{pmatrix}$$

With $\psi$ the spinor wave function $\begin{pmatrix} \varphi \\ \bar{\chi} \end{pmatrix} = \begin{pmatrix} \varphi_A \\ \bar{\chi}^{\dot{A}} \end{pmatrix}$ we have

$$\psi^c = C\bar{\psi}^T = C(\psi^+\gamma^0)^T = i\gamma^0\gamma^2\gamma^{0T}\psi^* = i\begin{pmatrix} 0 & -\sigma^2 \\ \sigma^2 & 0 \end{pmatrix}\begin{pmatrix} \varphi^* \\ \bar{\chi}^* \end{pmatrix} = i\begin{pmatrix} -\sigma^2\bar{\chi}^* \\ \sigma^2\varphi^* \end{pmatrix}$$

Now note that;

$$\varphi^* = (\varphi_A)^* = \bar{\varphi}_{\dot{A}} \text{ and } \chi = \chi^A = (\bar{\chi}^{\dot{A}})^*$$

Thus

$$\psi^c = i\begin{pmatrix} -\sigma^2\bar{\chi}^* \\ \sigma^2\varphi^* \end{pmatrix} = \begin{pmatrix} \begin{pmatrix} 0 & -1 \\ 1 & 0 \end{pmatrix}\bar{\chi}^* \\ \begin{pmatrix} 0 & 1 \\ -1 & 0 \end{pmatrix}\varphi^* \end{pmatrix} = \begin{pmatrix} \varepsilon_{AB}(\bar{\chi}^{\dot{A}})^* \\ \varepsilon^{\dot{A}\dot{B}}(\varphi_A)^* \end{pmatrix} = \begin{pmatrix} \varepsilon_{AB}\chi^A \\ \varepsilon^{\dot{A}\dot{B}}\bar{\varphi}_{\dot{A}} \end{pmatrix} = \begin{pmatrix} \chi_B \\ \bar{\varphi}^{\dot{B}} \end{pmatrix} \in H$$

where the $\varepsilon$ matrices are the spinor metric 2x2 matrices.

**The Mass Zero Case as a Graded Lie Algebra**

We start with the properties of a $Z_2$-graded Lie algebra $L = L_0 \oplus L_1$ where $L_0$ is the Lie algebra spanned by the generators $P_\mu$ ($\mu = 0,1,2,3$) and

$L_1$ is spanned by the spinor charge generators $Q_\alpha$ ($\alpha = 0,1,2,3$).

From the definition of a $Z_2$-graded Lie algebra $L$;

$P_\mu \in L_0, Q_\alpha \in L_1$ with gradings 0 and 1

$$(P_\mu, Q_\alpha) \to P_\mu \circ Q_\alpha = P_\mu Q_\alpha - (-1)^{0 \times 1} Q_\alpha P_\mu = P_\mu Q_\alpha - Q_\alpha P_\mu = [P_\mu, Q_\alpha] = 0$$

Similarly,

We also have $Q_\alpha \circ Q_\beta = Q_\alpha Q_\beta - (-1)^{1 \times 1} Q_\beta Q_\alpha = Q_\alpha Q_\beta + Q_\beta Q_\alpha = \{Q_\alpha, Q_\beta\}$, the anticommutator.

Since $Q_\alpha$ is a non-Hermitian operator (by construction), we can also consider the complex conjugate Dirac 4-spinor $Q_\beta \in \dot{F}$. To resolve differences in the literature we assume $\bar{Q}_\beta = \bar{Q}_{\dot{\beta}}$.

To constrain the number of options it is convenient at this stage to assume;

$Q = Q^c = C\bar{Q}^T$

Thus $(CQ)^T = (C^2\bar{Q}^T)^T = -\bar{Q}$ since $C^2 = -1$.

Hence $Q^T C = \bar{Q}$ since $C^T = -C$

Following now the logic of [8] in general, we consider from the $Z_2$ grading,

$\{Q_\alpha, Q_\beta\} = Q_\alpha Q_\beta + Q_\beta Q_\alpha = a(\gamma^\mu C)_{\alpha\beta} P_\mu$

Multiplying from the right by $C$; $Q_\alpha Q_\beta C_{\beta\delta} + Q_\beta C_{\beta\delta} Q_\alpha = a(\gamma^\mu)_{\alpha\gamma} C_{\gamma\beta} C_{\beta\delta} P_\mu = -a(\gamma^\mu)_{\alpha\delta} P_\mu$

Hence $Q_\alpha (Q^T C)_\delta + (Q^T C)_\delta Q_\alpha = -a(\gamma^\mu)_{\alpha\delta} P_\mu$

Thus $\{Q_\alpha, \bar{Q}_\beta\} = -a(\gamma^\mu)_{\alpha\delta} P_\mu$

We assume the operators $Q, \bar{Q}$ are Marjorana 4-spinors; then in 2-spinor notation we can simply replace the Dirac $\gamma$ matrices with their equivalent Pauli matrices yielding the following relationship;

$\{Q_A, \bar{Q}_{\dot{B}}\} - a(\sigma^\mu)_{A\dot{B}} P_\mu$ ............................................................(2)

Similarly we can show that, for 4-spinors, $\{\bar{Q}_\alpha, \bar{Q}_\beta\} = -a(C^{-1}\gamma^\mu)_{\alpha\beta} P_\mu$

Hence in 2-spinor notation, $\{\bar{Q}_A, \bar{Q}_{\dot{B}}\} = -a(C^{-1}\sigma^\mu)_{A\dot{B}} P_\mu$

Since all of the latter equations are relativistically invariant, we can transform them to the rest frame where $P_\mu = (E, 0, 0, 0) = (m, 0, 0, 0)$ with the speed of light normalised at $c = 1$.

With these values of $P_\mu$ in equation (2) above we have

$\{Q_A, \bar{Q}_{\dot{B}}\} = -a\sigma^0_{A\dot{B}} P_0$

Hence $\{Q_A, \bar{Q}_{\dot{B}}\} \sigma^0_{\dot{B}A} = -a\sigma^0_{A\dot{B}} \sigma^0_{\dot{B}A} P_0 = -aP_0$

Taking $A = 1$, $\dot{B} = \dot{1}$ we have $\{Q_1, \bar{Q}_{\dot{1}}\} \sigma^0_{\dot{1}1} = -aP_0$

Similarly for $A = 2$, $\dot{B} = \dot{2}$ we have $\{Q_2, \bar{Q}_{\dot{2}}\} \sigma^0_{\dot{2}2} = -aP_0$

Since $\sigma^0_{\dot{1}1} = \sigma^0_{\dot{2}2} = 1$ we have $\{Q_1, \bar{Q}_{\dot{1}}\} + \{Q_2, \bar{Q}_{\dot{2}}\} = -2aP_0$

For consistency with the current literature, we assume the constant $a = -1$.

Thus we have the quantum operator equality $\{Q_1, \bar{Q}_{\dot{1}}\} + \{Q_2, \bar{Q}_{\dot{2}}\} = 2P_0$

The left hand side of this expression is a positive definite quantum operator thus for $\psi$ an element of the Hilbert space;

If $\psi$ is the vacuum state then $<\psi, P_0\psi> = 0$ is equivalent to $2<\psi, Q_1\bar{Q}_{\dot{1}}\psi> + 2<\psi, Q_2\bar{Q}_{\dot{2}}\psi> = 0$

Since, for 2-spinors, $\varphi_A^* = \bar{\varphi}_{\dot{A}}$ we can rewrite this as: $2<\psi, Q_1 Q_1^*\psi> + 2<\psi, Q_2 Q_2^*\psi> \geq 0$

Thus $2\|Q_1\psi\|^2 + 2\|Q_2\psi\|^2 = 0$, from which we deduce that $Q_1|\psi> = Q_2|\psi> = 0$

and also $Q_{\dot{1}}|\psi> = Q_{\dot{2}}|\psi> = 0 \Rightarrow \omega_\psi(Q_1) = \omega_\psi(Q_2) = \omega_\psi(Q_{\dot{1}}) = \omega_\psi(Q_{\dot{2}}) = =\omega_\psi(P_0) = 0$.

## Factorial Representations of the $Z_2$ graded Lie algebra

The extension of the space $L_0$ to the space $L = L_0 \oplus L_1$ maintains $P^2$ as an element of the centre;

$$[P^2, Q_A] = 0 = [P^2, \bar{Q}^{\dot{A}}]$$

In a factorial representation $\Pi$ of the $Z_2$ graded algebra, it follows that $\Pi(P^2) = m^2 I$, fixing the mass $m$ of all particles in this representation. However, the spins of the particles in this representation are not fixed at a common value.

In this factorial 2-spinor representation, we have the algebraic identity ;

$\{Q_A, \bar{Q}_{\dot{B}}\} = \sigma^\mu_{A\dot{B}} P_\mu$ from equation (A) with the parameter $a=-1$.

With $P^2 = m^2 I$ in this representation, and setting $P_\mu = (m, 0, 0, 0)^T$ in the rest frame, we have the following set of identities from equation (2) taking $A = 1, 2$ and $\dot{B} = \dot{1}, \dot{2}$;

$$\{Q_1, \bar{Q}_{\dot{1}}\} = \sigma^\mu_{1\dot{1}} P_\mu = \sigma^0_{1\dot{1}} m = m$$
$$\{Q_1, \bar{Q}_{\dot{2}}\} = \sigma^\mu_{1\dot{2}} P_\mu = \sigma^0_{1\dot{2}} m = 0$$
$$\{Q_A, Q_B\} = \{\bar{Q}_{\dot{A}}, \bar{Q}_{\dot{B}}\} = 0$$

From these properties we see that the 2-spinors form at least a Clifford algebra in this factorial representation, and we see also that $Q_A^2 = 0$ and this is in fact a Grassmann algebra.

If $|p, \lambda>$ is an eigenstate in the Hilbert space $H$, then $P_\mu|p, \lambda> = p_\mu|p, \lambda>$.

The corresponding (pure) vector state is invariant under the translational group since we have

$$0 = P_\mu Q_A|p, \lambda> = Q_A P_\mu|p, \lambda> = Q_A p_\mu|p, \lambda> = p_\mu Q_A|p, \lambda>$$

We can thus always choose a minimum energy pure vector state $\omega_{|p, \lambda>}$ which is translation invariant and with $\omega_{|p, \lambda>}(Q_A) = <p, \lambda|Q_A|p, \lambda> = 0$. It is specified by its mass-energy $p$ and its spin value $\lambda$. Thus $\omega_{|p, \lambda>}$ is a translation invariant ergodic pure state [4].

If we now, exploiting local special relativistic covariance, choose an inertial reference frame in which the Wigner little group contains the spin generating 2x2 rotation matrices in the x-y plane, we have;

$$P_\mu = (E, 0, 0, E) \text{ therefore;}$$
$$\{Q_A, \bar{Q}_{\dot{B}}\} | p, \lambda > = \sigma^\mu_{A\dot{B}} P_\mu | p, \lambda > = \sigma^\mu_{A\dot{B}} p_\mu | p, \lambda >$$
$$\text{Now } \sigma^\mu_{A\dot{B}} p_\mu = \sigma^0 p_0 - \sigma^1 p_1 - \sigma^2 p_2 - \sigma^3 p_3 = E(\sigma^0 - \sigma^3)$$

$$= E\left(\begin{pmatrix} 1 & 0 \\ 0 & 1 \end{pmatrix} - \begin{pmatrix} 1 & 0 \\ 0 & -1 \end{pmatrix}\right) = \begin{pmatrix} 0 & 0 \\ 0 & 2E \end{pmatrix}$$

Hence applying this to our translation invariant vacuum state, $\omega_{|p,2>}$ we have;

$|< p, 2 | Q_1 \bar{Q}_{\dot{1}} | p, 2 >= - < p, 2 | \bar{Q}_{\dot{1}} Q_1 | p, 2 >= 0$

Similarly, considering the $2\dot{1}$ element of the matrix, we have

$|< p, 2 | Q_2 \bar{Q}_{\dot{1}} | p, 2 >= - < p, 2 | \bar{Q}_{\dot{1}} Q_2 | p, 2 >= 0$

Thus $(aQ_1 + bQ_2)\bar{Q}_{\dot{1}} | p, 2 >= 0$ for any scalars $a$ and $b$.

Since $Q_1$ and $Q_2$ span the subspace $L_1$, it follows that $\bar{Q}_{\dot{1}} | p, 2 >= 0$.

We conclude that the translation invariant pure Boson state $\omega_{|p,2>}$ is the local Clifford vacuum, and a spin-2 pure state corresponding to the graviton. It is an extreme point of the closed convex hull of the state space and is thus an extreme point of the translation invariant states: it is an ergodic state [4]. Its Supersymmetric fermion partner is the Gravitino with spin $\frac{3}{2}$. Thus from the properties of factorial representations of the translational group which are the generators of local diffeomorphisms and gravity; a minimal form of Supersymmetry naturally emerges, which is compatible with the existence of both the mass zero graviton, and its gravitino partner.

In Conclusion;

We predict a massless graviton and its supersymmetric partner on the basis of our analysis.

**Extension of Certain Results from Normal to Generalised Quantum States**

For this section, [1, 3, 4] are general references.

We define $\alpha: g \to \alpha_g$ to be a representation of *T* as automorphisms of the local fibre algebra **A(x)** which we assume to be isomorphic to a von Neumann algebra with trivial centre acting on a separable Hilbert space. The set $\{v(g): g \to f \circ \alpha_g; f \text{ is a quantum state, and } g \in T\}$ is a continuous group of commuting transformations of a subset of the dual space **A(x)***. If *f* is any quantum state of the algebra then define $\mathcal{E}$ to be the weak* closed convex hull of the set $\{v(g)f; g \in T\}$. $\mathcal{E}$ is a weak* compact convex set and each $v(g): \mathcal{E} \to \mathcal{E}$; by the Markov-Kakutani fixed point theorem it follows that $\mathcal{E}$ has an invariant element.

The group representation $\alpha$ acts ergodically on **A(x)** if given a projection *E* in **A(x)**, $\alpha_g(E) = E \; \forall g \in T$ implies that *E*=0 or *E*=*I*. Conversely, a quantum state *f* of A(x) is α-invariant if $f(\alpha_g(A)) = f(A) \quad \forall A$ in **A(x).** If *f* is an extreme point of the set of invariant states, then *f* is defined to be a *T*-ergodic state. Note that the set of invariant states is a non-void weak* compact convex subset of the generalised quantum state space of **A(x)** and is thus generated by its extreme points. Since there is a non-trivial invariant state for the amenable group *T*, thus there is an extremal *T*- invariant state of **A(x)**.

The fibre algebra **A(x)** is a quantum operator algebra and thus has an identity operator *I*. If *f* is a quantum state of **A(x)** then by definition, *f (I)=1*. The support of *f* is the unique smallest projection *E* in **A(x)** such that *f(E) =1* ; ie *f (I-E) = 0.*. Thus if *f* is separating, then the support of *f* is *I*.

We also require the following: *E* is an α-invariant projection in **A(x)** if $\alpha_g(E) = E \quad \forall g \in T$.

It now follows that if *f* is separating, then it is an ergodic state if and only if the representation $\alpha$ acts ergodically on the algebra **A(x)**. For let $\pi$ be the Gelfand-Naimark-Segal (GNS) representation of **A(x)** induced by the state *f* on the Hilbert space *H(f)*. Since *f* is separating $E_f = I$. In this case $\pi$ is a *-isomorphism, and $\pi \circ f = \omega_\xi$

Consider now the involution mapping on π(**A(x)**) defined as $A \to A^*$. This induces an anti-linear mapping on a dense subset of the Hilbert space *H(f)*; $S: A\xi \to A^*\xi$. Moreover, this extends to a mapping with closed graph which we also denote by *S*. By the theorem of Tomita-Takesaki [9] $S$ has a polar decomposition $S = J\Delta^{\frac{1}{2}}$ such that $J\pi(A(x))J=\pi(A(x))$'; the commutant of the fibre algebra $\pi(A(x))$. If $x = B\xi$ is in the domain of $S$, then it follows that $U_g x$ also lies in the domain of $S$, and we have the relationship;

$$U_g SB\xi = U_g B^*\xi = \alpha_g(B^*)\xi = \alpha_g(B)^*\xi = S\alpha_g(B)\xi = SU_g B\xi$$

This leads to the conclusion that, on the domain of $S$, we have $U_g S = SU_g$.

Then we have;

$$S = U_g SU_g^* = U_g J\Delta^{\frac{1}{2}}U_g^* = U_g JU_g^* U_g \Delta^{\frac{1}{2}} U_g^*$$

By uniqueness of the polar decomposition, $J = U_g JU_g^*$; $J$ and $U_g$ commute for all $g \in T$. From this we deduce that;

$B \in \pi(A(x)) \cap \{U_g; g \in T\}'$ implies that $JBJU_g = U_g JBJ$ for all $g \in T$.

Thus $JBJ \in \{U_g; g \in T\}' \cap \pi(A(x))'$.

Conversely, if $C \in \{U_g; g \in T\}' \cap \pi(A(x))'$ then $C = JBJ$ for some $B \in \pi(A(x))$ and $JBJU_g = U_g JBJ$ implies $JBU_g J = JU_g BJ$ and thus $BU_g = U_g B$ so that $B \in \{U_g; g \in T\}'$.

We conclude that;

$$J\left\{\pi(A(x)) \cap \{U_g; g \in T\}'\right\}J = \pi(A(x))' \cap \{U_g; g \in T\}' \quad \ldots\ldots\ldots\ldots\ldots(3)$$

The automorphic representation $\alpha: g \to \alpha_g$ of $T$ acts ergodically if and only if $\pi(A(x)) \cap \{U_g; g \in T\}'$ is trivial, containing only the projections 0 and I and thus consisting of the set of complex multiples of $I$. From the reasoning above, and noting that $J^2 = 1$, it follows that the representation $\alpha: g \to \alpha_g$ of $T$ acts ergodically if and only if $\pi(A(x))' \cap \{U_g; g \in T\}'$ is also trivial.

**In Conclusion; We predict a massless graviton and its supersymmetric partner on the basis of our analysis.**